\documentclass[a4paper,11pt]{article}
\pdfoutput=1 

\usepackage{jcappub}
\usepackage{xcolor}
\usepackage[T1]{fontenc} 
\usepackage{bm}
\usepackage{epsfig}
\usepackage[lofdepth,lotdepth,caption=false]{subfig}
\usepackage{graphicx}

\usepackage{float}
\usepackage{amsmath,amssymb}
\usepackage{textcomp}
\usepackage{gensymb}
\usepackage[utf8]{inputenc}
\usepackage{setspace}
\usepackage{hyperref}
\usepackage[normalem]{ulem}
\usepackage{pdflscape}
\usepackage{multirow}

\DeclareFontFamily{U}{mathb}{}
\DeclareFontShape{U}{mathb}{m}{n}{
  <-5.5> mathb5
  <5.5-6.5> mathb6
  <6.5-7.5> mathb7
  <7.5-8.5> mathb8
  <8.5-9.5> mathb9
  <9.5-11.5> mathb10
  <11.5-> mathb12
}{}
\DeclareSymbolFont{mathb}{U}{mathb}{m}{n}
\DeclareMathSymbol{\sqcdot}{\mathbin}{mathb}{"0D}

\def\gq#1#2{\theta_{#1#2}^{\rm g}}

\begin{document}

\title{Violation of Equivalence Principle in Neutrino Sector: Probing the Extended Parameter Space}
\author{Arman Esmaili}
\emailAdd{arman@puc-rio.br}
\affiliation{Departamento de F\'isica, Pontif\'icia Universidade Cat\'olica do Rio de Janeiro, Rio de Janeiro 22452-970, Brazil}

\abstract{The oscillation of neutrino flavors, due to its interferometry nature, is extremely sensitive to the phase differences developing during the propagation of neutrinos. In this paper we investigate the effect of the Violation of Equivalence Principle (VEP) on the flavor oscillation probabilities of atmospheric and cosmic neutrinos observed at neutrino telescopes such as IceCube. Assuming a general parameterization of VEP, dubbed {\it extended} parameter space, we show that the synergy between the collected data of high energy atmospheric and cosmic neutrinos severely constrains the VEP parameters. Also, the projected sensitivity of IceCube-Gen2 to VEP parameters is discussed.}
\maketitle
\date{\today}


\section{\label{sec:intro}Introduction}

The construction of neutrino telescopes, IceCube~\cite{Ahrens:2002dv} at the south pole, ANTARES~\cite{Collaboration:2011nsa} at Mediterranean Sea and Baikal-GVD in the Baikal lake at Russia~\cite{Belolaptikov:1997ry}, dawned a new era in our understanding of high energy Universe through the neutrino messenger. The observation of astrophysical (galactic and/or extragalactic) neutrinos by these neutrino telescopes was the primary target which has been accomplished by IceCube~\cite{Aartsen:2013bka,Aartsen:2013jdh}. One of the challenges in the observation of cosmic neutrinos is an efficient rejection of background events mainly consisting of atmospheric neutrinos and muons hitting the detector with orders-of-magnitude higher rates than the cosmic neutrinos. Various techniques have been developed to reject either or both of these backgrounds, from devoting a part of the detector for the veto to limiting the data selection to upgoing  $\mu$-tracks (originating mainly from charged current interaction of $\nu_\mu$ and $\bar{\nu}_\mu$) which benefits from Earth's matter filtering out the atmospheric muons.

Both the signal and background events in neutrino telescopes, respectively the cosmic and atmospheric neutrinos, provide powerful handles to probe new physics scenarios; such as neutrino decay~\cite{Beacom:2002vi,Beacom:2003nh,Baerwald:2012kc,Bustamante:2016ciw,Denton:2018aml,Abdullahi:2020rge,Bhattacharya:2010xj}, pseudo-Dirac neutrinos~\cite{Esmaili:2009fk,Esmaili:2012ac,Brdar:2018tce,Ahn:2016hhq,Joshipura:2013yba,Ahn:2019jbm,Shoemaker:2015qul}, sterile neutrinos~\cite{Esmaili:2012nz,Esmaili:2018qzu,Miranda:2018buo,Rajpoot:2013dha,Esmaili:2013vza,Aartsen:2020fwb,Esmaili:2013cja,Razzaque:2012tp,Razzaque:2011ab}, large extra dimensions~\cite{Esmaili:2014esa}, non-standard neutrino interactions~\cite{Esmaili:2013fva,Salvado:2016uqu,Mocioiu:2014gua,Choubey:2014iia,Demidov:2019okm,Rasmussen:2017ert,Aartsen:2017xtt}, neutrino-dark matter interaction~\cite{Farzan:2021gbx,Farzan:2018pnk}, heavy dark matter decay~\cite{Esmaili:2013gha,Feldstein:2013kka,Esmaili:2014rma,Bhattacharya:2019ucd,Kachelriess:2018rty,Chianese:2016opp,Murase:2015gea,Dekker:2019gpe,Bhattacharya:2017jaw}, violation of Lorentz symmetry~\cite{Hooper:2005jp,Aartsen:2017ibm,Abbasi:2010kx}, and violation of equivalence principle~\cite{Esmaili:2014ota,Fiorillo:2020gsb,Minakata:1996nd} (clearly, this list is not complete). In this paper we investigate the last-mentioned scenario, that is violation of equivalence principle (VEP), and its signatures on atmospheric and cosmic neutrinos.     

Oscillation of neutrino flavors is extremely sensitive to tiny phase differences that can be developed during the propagation of neutrino states. Consequently, any new physics scenario predicting a phase difference in addition to the standard one originating from the neutrino masses can be probed by neutrino flavor oscillations. The energy-dependence of these additional phase differences can be different from the mass-induced phase difference which is inversely proportional to the energy of propagating neutrinos. In our case of interest, the VEP-induced phases are proportional to the neutrino energy~\cite{Gasperini:1988zf,Gasperini:1989rt,Minakata:1994kt} and thus the atmospheric neutrinos are advantageous in constraining VEP, either in the $\sim$ few hundreds of MeV to $\sim$ multi-GeV range observed by Super-KamioKande~\cite{Battistoni:2005gy,Foot:1997kk,Foot:1998vr,Fogli:1999fs,GonzalezGarcia:2004wg} or particularly in the higher energy range observed at neutrino telescopes~\cite{Esmaili:2014ota,Fiorillo:2020gsb,Morgan:2007dj,Abbasi:2009nfa,GonzalezGarcia:2005xw}. The state-of-the-art limit on VEP parameters has been derived in~\cite{Esmaili:2014ota}, by analyzing the atmospheric neutrino data collected by IceCube in its final stages of construction, and updated in~\cite{Fiorillo:2020gsb} considering the more recent data sets. These limits have been derived under the assumption that the coupling of neutrinos to gravity is diagonal in the mass basis. However, the gravitational coupling of neutrinos can be diagonal in an arbitrary basis, such that in addition to the usual Pontecorvo-Maki-Nakagawa-Sakata matrix $U_{\rm PMNS}$ which transforms the mass basis to flavor basis, an extra $3\times3$ unitary matrix $U^{\rm g}$ transforming the gravitational basis to flavor basis has to be introduced. It has been correctly argued in~\cite{Fiorillo:2020gsb} that when $U^{\rm g}=I$, that is when the flavor and gravitational bases are identical, atmospheric neutrino data cannot constrain the VEP parameters while in this case measurements of the flavor content of cosmic neutrinos can constrain the VEP parameters~\cite{Minakata:1996nd}. Motivated by this complementarity of atmospheric and cosmic neutrino data in the search for VEP, introduced in~\cite{Fiorillo:2020gsb}, in this paper we investigate the extent to which the {\it extended} parameter space of arbitrary $U^{\rm g}$ can be probed\footnote{For future constraints that can be derived on VEP parameters, in the extended parameter space, from DUNE see~\cite{Diaz:2020aax}.}. Using the IceCube $\mu$-track data\footnote{Available at \url{https://icecube.wisc.edu/science/data-releases/}} collected between May 2010 and May 2012~\cite{Aartsen:2015rwa} and the current constraints on the flavor content of cosmic neutrinos~\cite{Abbasi:2020zmr}, we derive bounds on the extended parameter space of VEP scenario. Also, the projected limits from IceCube-Gen2 upgrade~\cite{Aartsen:2020fgd} will be discussed.             

The paper is structured as follows: in section~\ref{sec:atm} we explain in detail the phenomenology of atmospheric neutrino oscillations in the presence of VEP in its extended parameter space. Section~\ref{sec:cosmic} is devoted to the signature of VEP in the flavor content of cosmic neutrinos. In section~\ref{sec:data} the analysis of IceCube's atmospheric neutrino data is described. The derived limits on VEP parameters from both atmospheric and cosmic neutrino data are presented in section~\ref{sec:cons}. Finally, in section~\ref{sec:conc} we discuss and draw our conclusions from the obtained limits.

\section{\label{sec:prem}Signatures of VEP on atmospheric and cosmic neutrino oscillations}

The effect of VEP on neutrino oscillation, in the case $U^{\rm g}=U_{\rm PMNS}$, and its implementation in the Schr\"odinger-like equation of flavor oscillation for atmospheric neutrinos has been discussed in~\cite{Esmaili:2014ota}. In section~\ref{sec:atm} we elaborate on the features of atmospheric neutrino oscillation in the presence of VEP in its extended parameter space of arbitrary $U^{\rm g}$. Also, the effect of VEP on the flavor oscillation of cosmic neutrinos, studied in~\cite{Minakata:1996nd} and discussed in~\cite{Fiorillo:2020gsb}, will be reviewed in section~\ref{sec:cosmic}. The atmospheric neutrino dataset and the analysis method used for constraining the VEP parameters is presented in section~\ref{sec:data}.   

\subsection{\label{sec:atm}Atmospheric neutrino oscillations in the presence of VEP}

By taking into account the possibility of gravitationally induced neutrino flavor oscillation, originating from VEP, three sets of neutrino eigenstates can be identified: {\it i}) mass eigenstates $\nu_i$, $i=1,2,3$, which can be defined as the basis that diagonalizes the charged lepton mass matrix; {\it ii}) flavor eigenstates $\nu_\alpha$, $\alpha=e,\mu,\tau$, which enter into the charged-current interaction term of Lagrangian with charged leptons; {\it iii}) gravitational eigenstates $\nu^{\rm g}_i$, $i=1,2,3$, which is basis that diagonalizes the coupling of neutrinos to gravitational field. The VEP is introduced via the non-equality of gravitational couplings of $\nu^{\rm g}_i$, $G_{N,i}=\gamma_i G_N$, where $G_N$ is the gravitational constant. In terms of the parameters $\gamma_i$, VEP means that the diagonal matrix ${\rm diag}(\gamma_1,\gamma_2,\gamma_3)$ is not proportional to the identity matrix. As always, in the oscillation of neutrinos flavors one can rotate away one of the diagonal elements, so the physical combinations quantifying the VEP are $\Delta\gamma_{21}=\gamma_2-\gamma_1$ and $\Delta\gamma_{31}=\gamma_3-\gamma_1$. The $3\times3$ unitary matrix relating the mass and flavor bases is the usual PMNS matrix\footnote{Since the effect of VEP can be at maximum subleading, the tiny perturbation induced by VEP (if exists within the allowed parameter space) does not alter the values of the elements of $U_{\rm PMNS}$ derived from the global analysis of neutrino oscillation data.}, $U_{\rm PMNS}$, parameterized by three mixing angles $(\theta_{12},\theta_{13},\theta_{23})$ and one CP-violating phase $\delta$. We use the best-fit values of these mixing parameters from the latest global analysis of oscillation data~\cite{Esteban:2020cvm}. A natural assumption, made in several previous publications, is the equality of mass and gravitational bases. However, from a phenomenological point of view, these bases can be different~\cite{Gasperini:1988zf,Gasperini:1989rt} and so the gravitational basis is related to flavor basis by an unitary matrix $U^{\rm g}$. The equality of mass and gravitational bases means $U^{\rm g}=U_{\rm PMNS}$; while in general the matrix element of $U^{\rm g}$ can take any value (subject to unitarity condition). Being a unitary matrix, $U^{\rm g}$ can be parametrized with three mixing angles $(\gq12,\gq13,\gq23)$ and six phases where four of them can be rotated away by field redefinitions. To keep the analysis of this letter under control, we assume all the phases of $U^{\rm g}$ are vanishing\footnote{This can be justified by the argument of~\cite{Minakata:1996nd} showing that the effect of the phases of $U^{\rm g}$ is small when either vacuum or gravity oscillation dominates.}, so $U^{\rm g}=R_{23}(\gq23)R_{13}(\gq13)R_{12}(\gq12)$ where $R_{ij}$ is the rotation matrix in the $ij$-plane. 

By taking the metric $g_{\mu\nu}=\eta_{\mu\nu}+h_{\mu\nu}(x)$ in weak field approximation, where $\eta_{\mu\nu}={\rm diag}(1,-1,-1,-1)$ is the Minkowski metric and $h_{\mu\nu}(x)=-2\gamma_i\phi(x)\delta_{\mu\nu}$, with $\phi(x)$ being the Newtonian gravitational potential, the Klein-Gordon equation gives the following modified dispersion relation between energy $E_i$, momentum $p$ and mass $m_i$ of neutrinos: $E_i = p(1+2\gamma_i\phi) + \frac{m_i^2}{2p}(1+4\gamma_i\phi)$. Therefore, the Schr\"odinger-like equation of flavor oscillation is
\begin{equation}\label{eq:sch}
i \frac{{\rm d}\nu_\alpha}{{\rm d}r} = \left[ \frac{1}{2p} U_{\rm PMNS}\, \mathbf{M}^2 \, U_{\rm PMNS}^\dagger + \mathbf{V}(r) + 2p\, U^{\rm g}\, \mathbf{G}\,U^{{\rm g}\dagger} \right]_{\alpha\beta}\nu_\beta~,
\end{equation}       
where 
\begin{equation}
\mathbf{M}^2 =    \begin{pmatrix}
      0 & 0 & 0 \\
      0 & \Delta m^2_{21} & 0 \\
      0 & 0 & \Delta m^2_{31} \\
   \end{pmatrix}\;,\;
\mathbf{V}(r) =    \begin{pmatrix}
      \sqrt{2}G_F n_e & 0 & 0 \\
      0 & 0 & 0 \\
      0 & 0 & 0 \\   
       \end{pmatrix}\;,\;
\mathbf{G} =  \phi(r)  \begin{pmatrix}
      0 & 0 & 0 \\
      0 & \Delta\gamma_{21} & 0 \\
      0 & 0 & \Delta\gamma_{31} \\   
       \end{pmatrix}~,
\end{equation}        
where $G_F$ is the Fermi constant, $\Delta m_{ij}^2=m_i^2-m_j^2$ are neutrino mass-squared differences and $n_e(r)$ is the electron number density profile of Earth's matter taken from PREM~\cite{Dziewonski:1981xy}. The $\phi(r)$ in matrix $\mathbf{G}$ is the Newtonian gravitational potential profile along the propagation path. It is well-known that the main contribution to $\phi(r)$ at the Earth originates from the huge mass overdensities of Great Attractor, at the distance $\sim40$~Mpc and at the direction of Norma constellation, and Shapley Attractor at the distance $\sim150$~Mpc in the same direction~\cite{Kocevski:2005pj,Kocevski:2005kr}. Both of these attractors are massive superclusters with estimated masses $\sim3\times10^{15}$ and $\sim10^{16}$ solar masses, respectively for the Great and Shapley attractors. Each of these attractors contribute by $\sim5\times10^{-6}$ to the gravitational potential, adding up to $\phi_\oplus\sim10^{-5}$ at the Earth. For the atmospheric neutrinos the $r$-dependence of potential can be safely neglected, and so the gravitational potential enters as a multiplicative constant factor, $\phi_\oplus$, in front of $\Delta\gamma_{ij}$. The equation for anti-neutrinos can be obtained from Eq.~(\ref{eq:sch}) by $\mathbf{V}(r)\to-\mathbf{V}(r)$ and $U_{\rm PMNS}\to U_{\rm PMNS}^\ast$. 

The effect of VEP-term in Eq.~(\ref{eq:sch}) is to change the effective mixing parameters. Depending on the matrix $U^{\rm g}$, the effective mixing angles and/or mass-squared differences can deviate from their values in the standard picture of neutrino sector. In the case of $U_{\rm PMNS}=U^{\rm g}$, the effective mixing angles are equal to $(\theta_{12},\theta_{13},\theta_{23})$ and just the effective mass-squared differences are different from $\Delta m_{21}^2$ and $\Delta m_{31}^2$ by $4p^2\phi\Delta\gamma_{21}$ and $4p^2\phi\Delta\gamma_{31}$, respectively. The phenomenology of this case, including the shifts in the minima and maxima of atmospheric oscillation probabilities and the emergence of new resonances, has been discussed in detail in~\cite{Esmaili:2014ota}. In the case of $U_{\rm PMNS}\neq U^{\rm g}$, which is the main subject of this paper, not only the mass-squared differences will change, but also the effective mixing angles differ from their standard values. These changes induce new resonances and oscillation patterns that will be discussed in the following subsections. However, the general case of three non-zero and arbitrary $\theta^{\rm g}_{ij}$ is computationally difficult to probe, and thus, we restrict our analysis to the cases where one of the $\theta^{\rm g}_{ij}$ is non-zero at a time. Although for the purpose of deriving bounds on VEP parameters in section~\ref{sec:cons} we numerically solve Eq.~(\ref{eq:sch}) for the atmospheric neutrinos traversing the Earth to compute the oscillation probabilities $P_{\nu_\mu(\bar{\nu}_\mu)\to\nu_\mu(\bar{\nu}_\mu)}$ and $P_{\nu_\mu(\bar{\nu}_\mu)\to\nu_\tau(\bar{\nu}_\tau)}$ with zenith angles $-1\leq\cos\theta_z\leq0$ and energies $E_\nu\gtrsim100$~GeV, in the following we comment on the expected features for each $\theta^{\rm g}_{ij}\neq0$.  

\subsubsection{\label{sec:gq12}The case of $\gq12\neq0$}

\begin{figure}[t!]
\centering
\subfloat{
\includegraphics[width=0.49\textwidth]{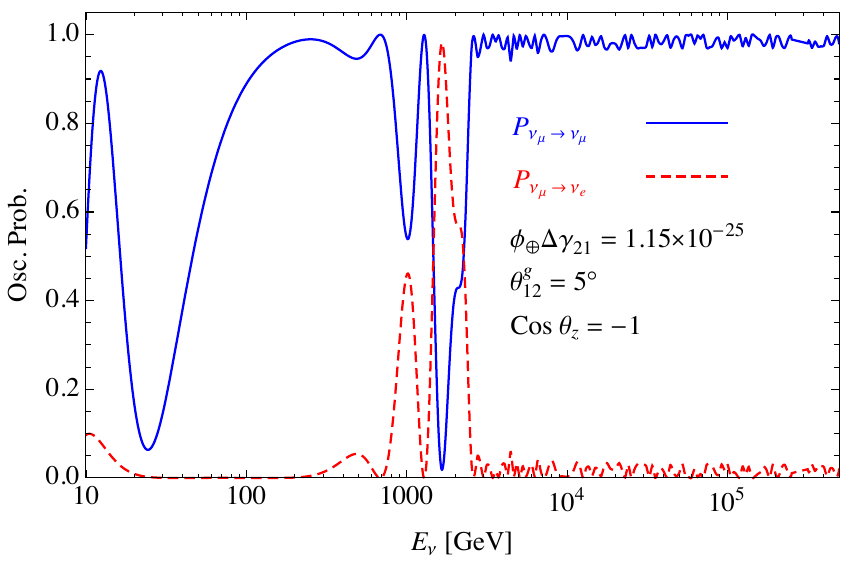}
\label{fig:nu-gq5-dg12}
}
\subfloat{
\includegraphics[width=0.49\textwidth]{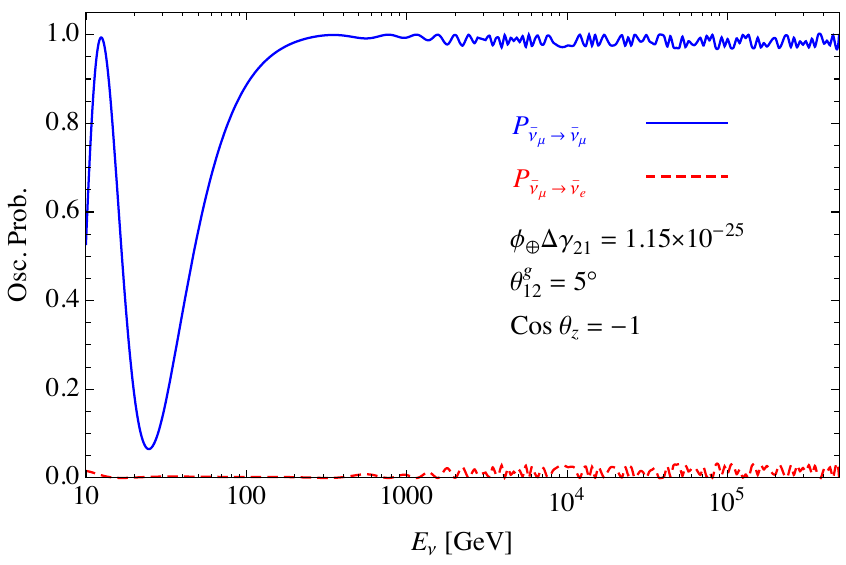}
\label{fig:antinu-gq5-dg12}
}
\quad
\subfloat{
\includegraphics[width=0.49\textwidth]{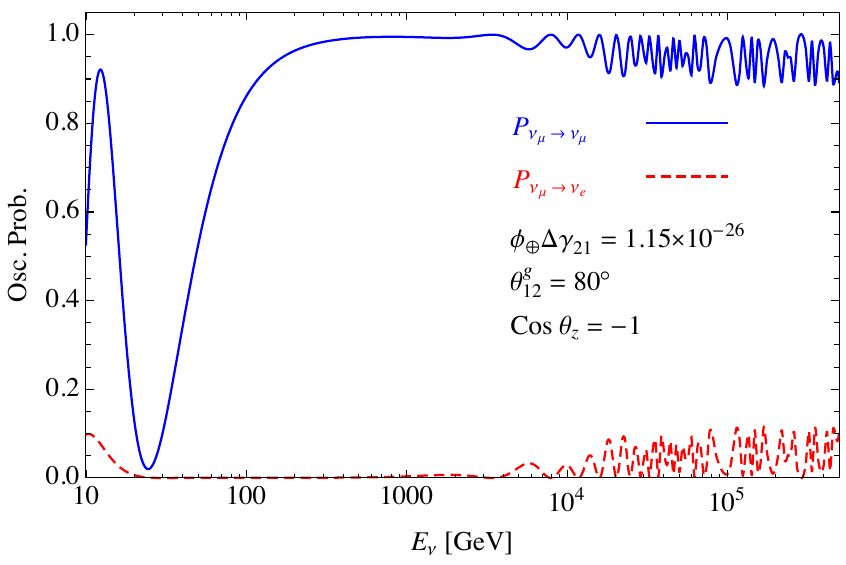}
\label{fig:nu-gq80-dg12}
}
\subfloat{
\includegraphics[width=0.49\textwidth]{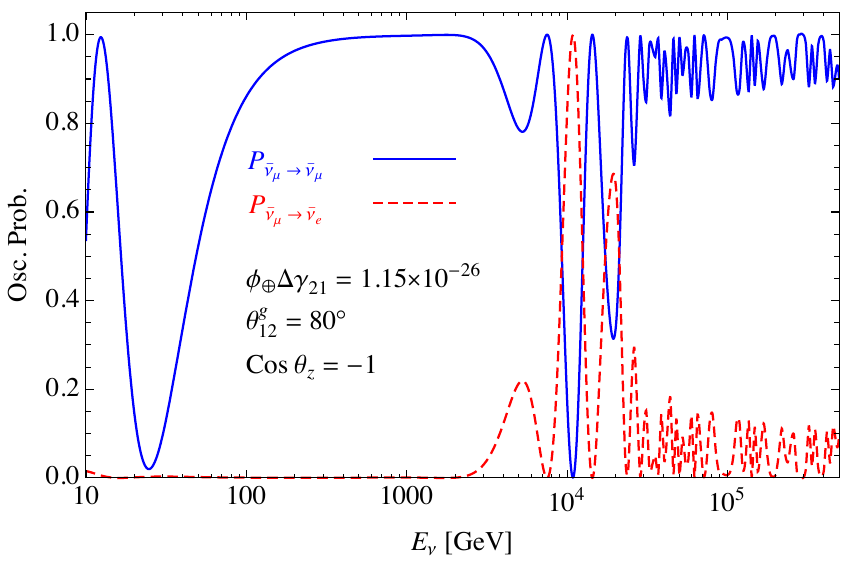}
\label{fig:antinu-gq80-dg12}
}
\caption{\label{fig:gq12}Oscillation probabilities $P_{\nu_\mu\to\nu_\mu(\nu_e)}$ (left panels) and $P_{\bar{\nu}_\mu\to\bar{\nu}_\mu(\bar{\nu}_e)}$ (right panels), for $(\gq12,\phi_\oplus\Delta\gamma_{21})=(5^\circ,1.15\times10^{-25})$ (top panels) and $(\gq12,\phi_\oplus\Delta\gamma_{21})=(80^\circ,1.15\times10^{-26})$ (bottom panels). For all the panels $\cos\theta_z=-1$. }
\end{figure}

Let us start with $\gq12\neq0$. In the high energy range, $E_\nu\gtrsim100$~GeV, the solar mass-squared difference $\Delta m_{21}^2$ and the mixing angles $\theta_{12}$ and $\theta_{13}$ can be neglected. The oscillation length induced by $\Delta m_{31}^2$ also will be larger than the diameter of Earth for these energies and so we have $P_{\nu_\mu\to\nu_\mu}=1$ and $P_{\nu_\mu\to\nu_\tau}=0$ in the absence of VEP. The VEP-term $2pU^{\rm g}\mathbf{G}U^{{\rm g}\dagger}$ takes the following block-diagonal form (replacing the momentum $p$ by $E_\nu$)
\begin{equation}\label{eq:gq12mat}
2E_\nu U^{\rm g}\mathbf{G}U^{{\rm g}\dagger} |_{{\rm just}\,\gq12\neq0} =  2E_\nu\phi_\oplus \begin{pmatrix}  \Delta\gamma_{21}\sin^2\gq12 & \Delta\gamma_{21}\sin\gq12\cos\gq12 & 0 \\
      \Delta\gamma_{21}\sin\gq12\cos\gq12 & \Delta\gamma_{21}\cos^2\gq12 & 0 \\
      0 & 0 & \Delta\gamma_{31} \\   
       \end{pmatrix}~.
\end{equation}
Clearly, $\Delta\gamma_{31}$ does not affect the oscillation pattern and so for the case of just $\gq12\neq0$, atmospheric neutrino data cannot constrain $\Delta\gamma_{31}$. In the high energy the $\nu_\tau$ decouples from other flavors, and the $e-\mu$ sector can be described by the following $2\times2$ Hamiltonian
\begin{equation}\label{eq:Hgq12}
H_{2\nu} = \begin{pmatrix}
      \sqrt{2}G_Fn_e & 0 \\
      0 & 0 \\
      \end{pmatrix}
      + 2E_\nu\phi_\oplus R_{12}({\gq12}) \begin{pmatrix}
      0 & 0 \\
      0 & \Delta\gamma_{21} \\
      \end{pmatrix} R_{12}^\dagger({\gq12})~.
\end{equation} 
From this Hamiltonian, the difference of instantaneous eigenvalues $\Delta H_m$ and the mixing angle $\Theta_m$ are
\begin{equation}
\Delta H_m = \sqrt{2}G_F n_e R \qquad, \qquad \sin^22\Theta_m = \frac{R_0^2\sin^22\gq12}{R^2}~,
\end{equation}  
where 
\begin{equation}\label{eq:R}
R= \left[ R_0 + \cos2\gq12\right]^2 + \sin^22\gq12 \qquad {\rm and} \qquad R_0 = \frac{-2E_\nu\phi_\oplus\Delta\gamma_{21}}{\sqrt{2}G_Fn_e}~.
\end{equation}
$R_0$ is the relative strength of VEP-induced and matter contributions, and takes the value 
\begin{equation}
R_0 \simeq -0.66 \left( \frac{E_\nu}{{\rm TeV}} \right) \left( \frac{\phi_\oplus\Delta\gamma_{21}}{10^{-25}} \right) \left( \frac{4\, N_{\rm A} {\rm cm}^{-3}}{\bar{n}_e} \right)~,
\end{equation}
where $N_{\rm A}$ is the Avogadro's number and $\bar{n}_e$ is the average electron number density along the propagation path, which is $\bar{n}_e\simeq4.12 N_{\rm A} {\rm cm}^{-3}$ for $\cos\theta_z=-1$. In the approximation of constant electron number density $\bar{n}_e$, the oscillation in $e-\mu$ sector over the propagation distance $L$, with the oscillation half-phase $\Delta H_mL/2$, can be casted as
\begin{equation}\label{eq:probgq12} 
P_{\nu_\mu(\bar{\nu}_\mu)\to\nu_e(\bar{\nu}_e)} = \sin^22\Theta_m \sin^2\left( \frac{\sqrt{2}G_Fn_eR}{2}\,L\right)~.
\end{equation}
The energy dependence of oscillation phase is through the resonance factor $R$ which increases with the increase in energy. At high energies, when $R_0\gg1$, there is a fast oscillation between $e-\mu$ flavors (for both neutrinos and anti-neutrinos). From Eq.~(\ref{eq:R}), there is a resonance when the resonance factor $R$ takes its minimum value, that is when $R_0 = -\cos2\gq12$. When $0<\gq12<\pi/4$, for $\Delta\gamma_{21} >0\, (<0)$ the resonance occurs for neutrinos (anti-neutrinos). Due to the sign change of $\cos2\gq12$, the opposite happens for $\pi/4<\gq12<\pi/2$: for $\Delta\gamma_{21} >0\, (<0)$ the resonance occurs for anti-neutrinos (neutrinos). From the resonance condition, the resonance energy can be derived as
\begin{equation}\label{eq:resgq12}
E_{\nu}^{{\rm res},\gq12} = \frac{\sqrt{2}G_Fn_e\cos2\gq12}{2\phi_\oplus\Delta\gamma_{21}}\simeq 1.5\,{\rm TeV} \left( \frac{\bar{n}_e}{4\, N_{\rm A} {\rm cm}^{-3}} \right) \left( \frac{10^{-25}}{\phi_\oplus\Delta\gamma_{21}} \right) \left( \frac{\cos2\gq12}{1} \right)~.
\end{equation} 
At the resonance we have $\sin^22\Theta_m = \cos^22\gq12$. For $\gq12=\pi/4$ we have $\sin^22\Theta_m =0$ and so the mixing disappears at the resonance; while, however, the fast oscillations with large mixing angle are present right above the resonance energy. To visualize these effects, we show in figure~\ref{fig:gq12} the oscillation probabilities $P_{\nu_\mu(\bar{\nu}_\mu)\to\nu_e(\bar{\nu}_e)}$ and $P_{\nu_\mu(\bar{\nu}_\mu)\to\nu_\mu(\bar{\nu}_\mu)}$ obtained from numerical solution of Eq.~(\ref{eq:sch}) for $\gq12\neq0$ and $\cos\theta_z=-1$. For $E_\nu\gtrsim100$~GeV, $P_{\nu_\mu(\bar{\nu}_\mu)\to\nu_e(\bar{\nu}_e)}=0$ and $P_{\nu_\mu(\bar{\nu}_\mu)\to\nu_\mu(\bar{\nu}_\mu)}=1$ in the standard oscillation scenario. In all the panels, the minimum in $\nu_\mu$ and $\bar{\nu}_\mu$ survival probabilities at $E_\nu\simeq24$~GeV originates from the standard oscillation in $\mu-\tau$ sector. The upper panels (the left panel for neutrinos and the right panel for anti-neutrinos) are for $(\gq12,\phi_\oplus\Delta\gamma_{21})=(5^\circ,1.15\times10^{-25})$. The resonance can be seen in the neutrino channel at the expected energy from Eq.~(\ref{eq:resgq12}). In both the neutrino and anti-neutrino channels, fast and small amplitude (due to the smallness of $\gq12$) oscillations can be identified. In the lower panels we chose $(\gq12,\phi_\oplus\Delta\gamma_{21})=(80^\circ,1.15\times10^{-26})$. Since $\gq12$ is in the second octant, the resonance moves to anti-neutrino channel and to a higher energy since $\phi_\oplus\Delta\gamma_{21}$ is smaller compared to top panels. Since the deviation of $\gq12$ from 0 or $\pi/2$ is larger in the lower panels, the amplitude of fast oscillations at high energies is larger.

\subsubsection{\label{sec:gq13}The case of $\gq13\neq0$}

\begin{figure}[t!]
\centering
\subfloat{
\includegraphics[width=0.49\textwidth]{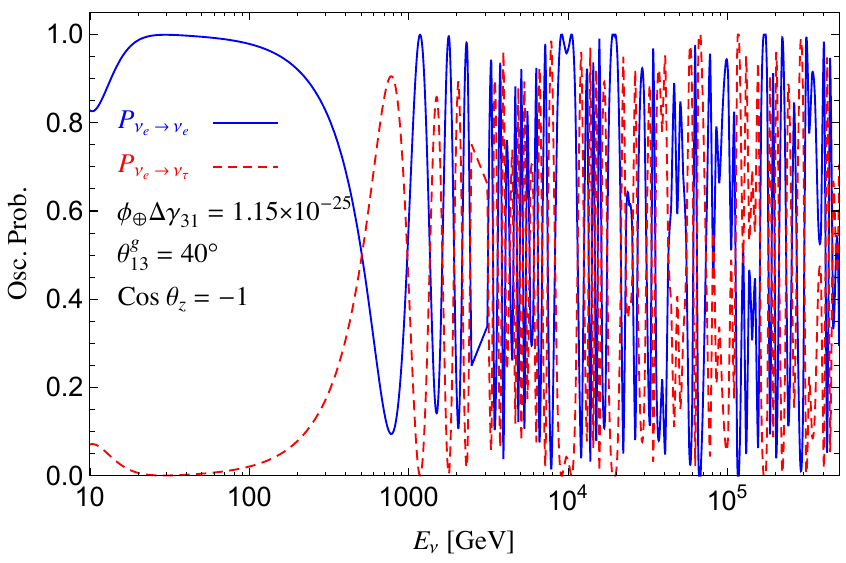}
\label{fig:nu-gq40-dg13}
}
\subfloat{
\includegraphics[width=0.49\textwidth]{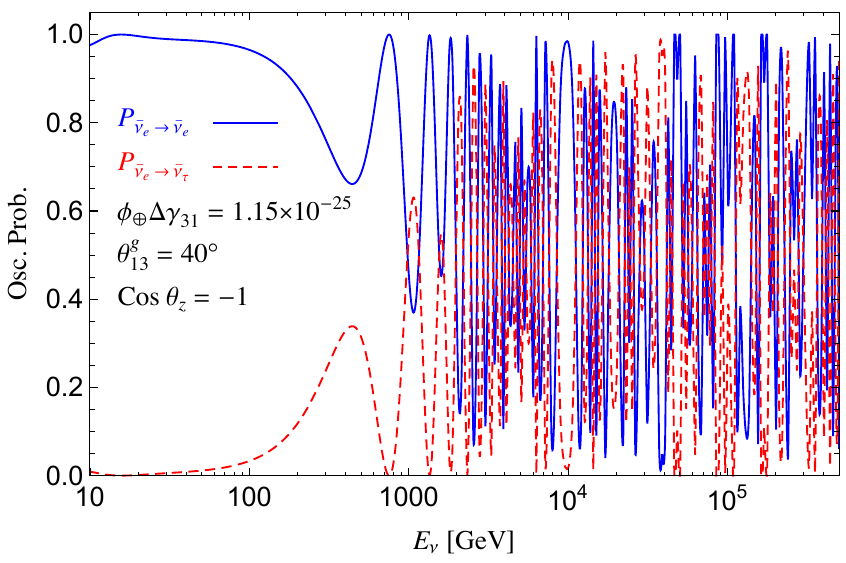}
\label{fig:antinu-gq40-dg13}
}
\caption{\label{fig:gq13}Oscillation probabilities $P_{\nu_e\to\nu_e(\nu_\tau)}$ (left panel) and $P_{\bar{\nu}_e\to\bar{\nu}_e(\bar{\nu}_\tau)}$ (right panel), for $(\gq13,\phi_\oplus\Delta\gamma_{31})=(40^\circ,1.15\times10^{-25})$. For both panels $\cos\theta_z=-1$.}
\end{figure}

For $\gq13\neq0$ the VEP-term takes the following form 
\begin{equation}\label{eq:gq13mat}
2E_\nu U^{\rm g}\mathbf{G}U^{{\rm g}\dagger} |_{{\rm just}\,\gq13\neq0} =  2E_\nu\phi_\oplus \begin{pmatrix}
      \Delta\gamma_{31}\sin^2\gq13 & 0 & \Delta\gamma_{31}\sin\gq13\cos\gq13 \\
      0 & \Delta\gamma_{21} & 0 \\
      \Delta\gamma_{31}\sin\gq13\cos\gq13 & 0 & \Delta\gamma_{31}\cos^2\gq13 \\   
       \end{pmatrix}~.
\end{equation}
In this case, $\Delta\gamma_{21}$ does not affect the atmospheric neutrino oscillation probabilities. The oscillation features are exactly the same as in section~\ref{sec:gq12}, replacing $(\gq12,\Delta\gamma_{21})$ by $(\gq13,\Delta\gamma_{31})$, but now in the $e-\tau$ sector. The resonance energy in the ${\nu_e(\bar{\nu}_e)\to\nu_\tau(\bar{\nu}_\tau)}$ and ${\nu_\tau(\bar{\nu}_\tau)\to\nu_\tau(\bar{\nu}_\tau)}$ oscillations is given by Eq.~(\ref{eq:resgq12}) with the replacement $\gq12\to\gq13$. Figure~\ref{fig:gq13} shows the oscillation probabilities $P_{\nu_e\to\nu_e(\nu_\tau)}$ (left panel) and $P_{\bar{\nu}_e\to\bar{\nu}_e(\bar{\nu}_\tau)}$ (right panel) for $(\gq13,\phi_\oplus\Delta\gamma_{31})=(40^\circ,1.15\times10^{-25})$ for up-going neutrinos at IceCube ($\cos\theta_z=-1$).

\subsubsection{\label{sec:gq23}The case of $\gq23\neq0$}

In the case of $\gq23\neq0$, both non-vanishing $\Delta\gamma_{21}$ and $\Delta\gamma_{31}$ alter the oscillation probabilities of atmospheric neutrinos. The VEP-term in this case is 
\begin{equation}\label{eq:gq13mat}
2E_\nu U^{\rm g}\mathbf{G}U^{{\rm g}\dagger} |_{{\rm just}\,\gq23\neq0} =  2E_\nu\phi_\oplus \begin{pmatrix}
      0 & 0 & 0 \\
      0 & \Delta\gamma_{21} \cos^2\gq23+\Delta\gamma_{31} \sin^2\gq23 & \frac{1}{2} (\Delta\gamma_{31}-\Delta\gamma_{21})\sin 2\gq23 \\
      0 & \frac{1}{2} (\Delta\gamma_{31}-\Delta\gamma_{21})\sin 2\gq23 & \Delta\gamma_{31} \cos^2\gq23+\Delta\gamma_{21} \sin^2\gq23 \\ 
       \end{pmatrix}~.
\end{equation} 
In the high energy, the oscillation occurs in the $\mu-\tau$ sector and the $\nu_e(\bar{\nu}_e)$ decouples. The VEP-term is diagonal when $\Delta\gamma_{31}=\Delta\gamma_{21}$, so in this case the oscillation probabilities do not deviate from standard scenario. Since the matter effect $\mathbf{V}(r)$ decouples, there is no resonant flavor conversion in the $\mu-\tau$ sector. The oscillation pattern in high energy ($\gtrsim100$~GeV) for $\gq23\neq0$ can be described by an effective mass-squared difference $\Delta m^2_{\rm eff}=2E_\nu^2\phi_\oplus(\Delta\gamma_{31}-\Delta\gamma_{21})=2E_\nu^2\phi_\oplus\Delta\gamma_{32}$ and the mixing angle $\gq23$ (see~\cite{Esmaili:2014ota} for more details). The oscillation probabilities for neutrinos and anti-neutrinos are the same. 

\begin{figure}[t!]
\centering
\subfloat{
\includegraphics[width=0.49\textwidth]{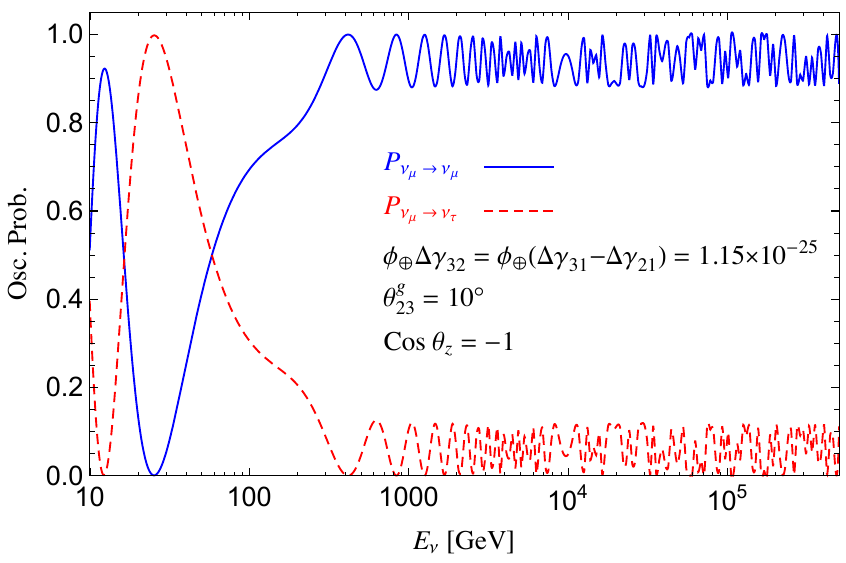}
\label{fig:nu-gq40-dg13}
}
\subfloat{
\includegraphics[width=0.49\textwidth]{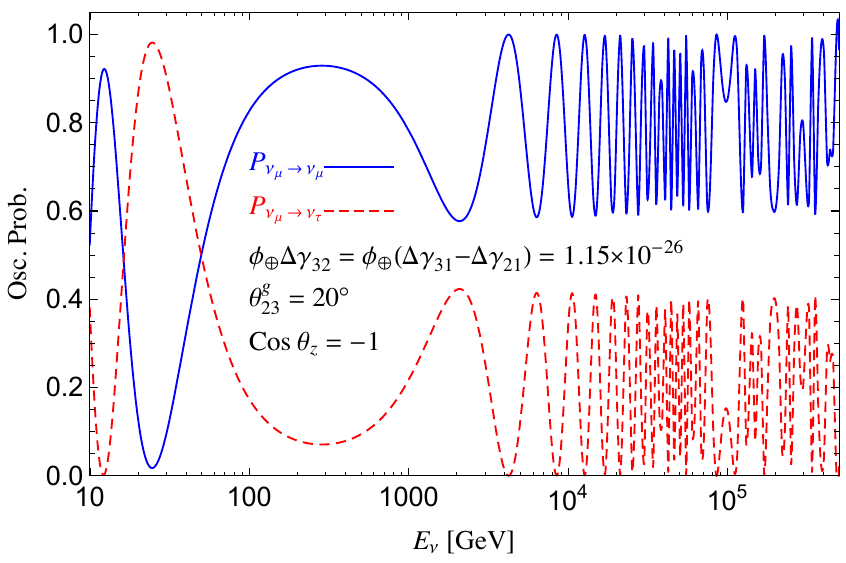}
\label{fig:antinu-gq40-dg13}
}
\caption{\label{fig:gq23}Oscillation probabilities $P_{\nu_\mu\to\nu_\mu(\nu_\tau)}$ for $(\gq23,\phi_\oplus\Delta\gamma_{32})=(10^\circ,1.15\times10^{-25})$ (left panel) and $(\gq23,\phi_\oplus\Delta\gamma_{32})=(20^\circ,1.15\times10^{-26})$ (right panel). For both panels $\cos\theta_z=-1$. In the high energy, $\gtrsim100$~GeV, the oscillation probabilities for anti-neutrinos, $P_{\bar{\nu}_\mu\to\bar{\nu}_\mu(\bar{\nu}_\tau)}$, are the same as neutrinos.}
\end{figure}

Figure~\ref{fig:gq23} shows the oscillation probabilities $P_{\nu_\mu\to\nu_\mu(\nu_\tau)}$ for two examples of VEP parameters: $(\gq23,\phi_\oplus\Delta\gamma_{32})=(10^\circ,1.15\times10^{-25})$ in the left panel and $(\gq23,\phi_\oplus\Delta\gamma_{32})=(20^\circ,1.15\times10^{-26})$ in the right panel. Oscillation probabilities for anti-neutrinos are the same. The high energy fast oscillations originate from the ``vacuum''-like oscillation induced by $\Delta m^2_{\rm eff}=2E_\nu^2\phi_\oplus\Delta\gamma_{32}$ with the amplitude moderated by $\gq23$ which is larger in the right panel compared to the left panel. A decrease in $\Delta\gamma_{32}$ leads to a shift of fast oscillations to higher energies, as can be seen in the right panel.       

\subsection{\label{sec:cosmic}Oscillation of cosmic neutrino flavors in the presence of VEP}

The VEP can change the flavor content of cosmic neutrinos observed by IceCube~\cite{Minakata:1996nd,Fiorillo:2020gsb}. It is well-known that the cosmic neutrinos, due to the large propagation distances, experience decoherent flavor oscillation. In the standard scenario, that is without VEP, the oscillations change the flavor ratio of neutrino flux at the source $(r_e:r_\mu:r_\tau)|_{\rm S}$ to $(r_e:r_\mu:r_\tau)|_\oplus$ at the Earth, where $r_\alpha |_\oplus = \sum_\beta P_{\alpha\beta}\, r_\beta |_{\rm S}$ and $P_{\alpha\beta} = \sum_i |(U_{\rm PMNS})_{\alpha i}|^2 |(U_{\rm PMNS})_{\beta i}|^2$. Here $r_\alpha$ denotes the sum of $\nu_\alpha$ and $\bar{\nu}_\alpha$ flavors. The source flavor ratio depends on the production mechanism of cosmic neutrinos, commonly assumed through the pion decay chain $\pi^\pm\to\mu^\pm + \nu_\mu(\bar{\nu}_\mu)\to e^\pm+\nu_e(\bar{\nu}_e)+\bar{\nu}_\mu+\nu_\mu$ which leads to $(r_e:r_\mu:r_\tau)|_{\rm S}\simeq1:2:0$~\cite{Lipari:2007su}. The muons from the pion decay can experience substantial energy-loss before the decay by virtue of interactions with the ambient gas and/or the magnetic field in the production site. In this case, known as the damped muon scenario, the flavor ratio at the source is $\simeq 0:1:0$. With the current best-fit values of the elements of $U_{\rm PMNS}$ from~\cite{Esteban:2020cvm}, the decoherent oscillations change the source flavor ratio $(r_e:r_\mu:r_\tau)|_{\rm S}=1/3:2/3:0$ to $(r_e:r_\mu:r_\tau)|_\oplus=0.30 : 0.36 : 0.34$ and $(r_e:r_\mu:r_\tau)|_{\rm S}=0:1:0$ to $(r_e:r_\mu:r_\tau)|_\oplus=0.17 : 0.45 : 0.37$.   

In the presence of VEP the averaged oscillation probability $P_{\alpha\beta}$ will change. The oscillation probabilities can still be calculated from Eq.~(\ref{eq:sch}) after appropriate averaging due to large propagation distances. The matter potential experienced by cosmic neutrinos is completely negligible~\cite{Minakata:1996nd} and so the $\mathbf{V}(r)$ term can be dropped. To simplify the discussion let us consider the $2\nu$ approximation which consists of one mass-squared difference $\Delta m^2$ and (vacuum) mixing angle $\theta$, while the VEP term contains one $\Delta\gamma$ and one mixing angle $\theta^{\rm g}$. It is straightforward to verify that the total $2\nu$ Hamiltonian can be written in terms of an effective mixing angle $\theta_{\rm eff}$ given by
\begin{equation}\label{eq:effangle}
\tan2\theta_{\rm eff} = \frac{\Delta m^2\sin2\theta+4E_\nu^2\phi_{\rm IG}\Delta\gamma\sin2\theta^{\rm g}}{\Delta m^2\cos2\theta+4E_\nu^2\phi_{\rm IG}\Delta\gamma\cos2\theta^{\rm g}}~,
\end{equation}
where $\phi_{\rm IG}$ is the gravitational potential in intergalactic space. Although cosmic neutrinos experience also the Galactic and ``in site'' gravitational potentials, it has been shown in~\cite{Minakata:1996nd} that their contributions are smaller than $\phi_{\rm IG}$ and can be neglected. The value of $\phi_{\rm IG}$ and its profile considerably depends on the distance and direction of production site. For simplification, justified by the limit we are going to consider where the VEP-induced oscillation predominates over the mass-induced one, we assume a constant $\phi_{\rm IG}$. The value of intergalactic gravitational potential would be $\simeq10^{-5}$ for sources as far as $\simeq100$~Mpc for directions in the half-sky containing the Great and Shapley Attractors, while for other directions smaller values would be expected. The phenomenology of the effective mixing angle in Eq.~(\ref{eq:effangle}), including the resonances induced by gravitational potential, has been discussed in detail in~\cite{Minakata:1996nd}. From Eq.~(\ref{eq:effangle}), when $4E_\nu^2\phi_{\rm IG}\Delta\gamma\gg\Delta m^2$ the VEP-induced oscillation dominates and $\theta_{\rm eff}\to\theta^{\rm g}$. Numerically, this condition implies
\begin{equation}\label{eq:vepdom}
\phi_{\rm IG}\Delta\gamma \gg 6.2\times10^{-32} \left( \frac{\Delta m^2}{2.5\times10^{-3}~{\rm eV}^2} \right) \left( \frac{100~{\rm TeV}}{E_\nu} \right)^2~, 
\end{equation}
where $E_\nu\simeq100$~TeV is the average energy of cosmic neutrinos observed by IceCube and we chose the atmospheric mass-squared difference for the scale of $\Delta m^2$. Thus, when the condition in Eq.~(\ref{eq:vepdom}) is satisfied, the oscillation of cosmic neutrinos is entirely governed by $\theta_{ij}^{\rm g}$ and we have $P_{\alpha\beta} = \sum_i |U^{\rm g}_{\alpha i}|^2 |U^{\rm g}_{\beta i}|^2$. Consequently, IceCube's measurement of the flavor ratio of cosmic neutrinos can place limits on the elements of $U^{\rm g}$.

\subsection{\label{sec:data}IceCube's data and analysis method}

We use the 2-years IceCube's data set of $\mu$-tracks~\cite{Aartsen:2015rwa} from Northern sky collected from May 2010 to May 2012\footnote{Obviously, IceCube has collected more data during the last $\sim10$ years, which can be used for constraining the VEP. However, the constraints derived from all these data sets, including the one we use here, are statistically saturated. We use the data set that has a complete published detector information (simulation, systematic errors, etc).}. This data set consists of 35,322 $\mu$-track events with $-1\leq\cos\theta_z\leq0.1$ and $21.98~{\rm GeV}\leq E_\mu^{\rm proxy}\leq 289.9~{\rm TeV}$. This data set provided the first evidence of astrophysical neutrinos in $\mu$-tracks, at $3.7\sigma$ C.L., discovering 21 events with signal probabilities $>33\%$. For our purpose of studying the energy and zenith distributions of atmospheric neutrinos, the $\lesssim 0.05\%$ contamination from cosmic neutrinos can be safely neglected.   

The $\nu_\mu$, $\bar{\nu}_\mu$, $\nu_\tau$ and $\bar{\nu}_\tau$ effective areas and their uncertainties are published in the data release accompanying~\cite{Aartsen:2015rwa}, where the last two take into account the $\tau^\pm$ production in the charged-current interaction of $\nu_\tau$ and $\bar{\nu}_\tau$ with the subsequent leptonic decay of $\tau^\pm$ which produce $\mu^\pm$. The published effective areas are three dimensional tables providing the MC simulation of detector (separately for 2010 and 2011 configurations) in bins of true neutrino energy $E_\nu$, reconstructed muon energy proxy $E_\mu^{\rm proxy}$ and reconstructed zenith angle $\theta_z$ which is assumed to be the same as true zenith angle within the binning. The number of bins for $E_\nu$, $E_\mu^{\rm proxy}$ and $\cos\theta_z$ are, respectively, 280, 50 and 11; covering the ranges $10^2\leq E_\nu/{\rm GeV} \leq10^9$, $10^2\leq E_\mu^{\rm proxy}/{\rm GeV} \leq10^7$ and $-1\leq \cos\theta_z \leq0.1$.  We will denote the $\nu_\alpha$ effective area in the $(i,j,k)$th bins of $(E_\nu, E_\mu^{\rm proxy},\cos\theta_z)$ by $A_{ijk}^{\nu_\alpha}$. To compute the expected number of events in the $(j,k)$th bins of observable quantities, $E_\mu^{\rm proxy}$ and $\cos\theta_z$, the effective areas should be convoluted by the corresponding averaged atmospheric neutrino flux in these bins. We use the bin-averaged atmospheric neutrino fluxes $\Phi^{\nu_\mu}_{ik}$ provided by the IceCube, based on~\cite{Honda:2004yz}, that have been used in the analysis of this data set and include also the corrections for the simulated detector optical efficiency, $\eta_{ijk}$, that should be multiplied by the effective areas. With these information, the expected number of events in the $(j,k)$th bins of $(E_\mu^{\rm proxy},\cos\theta_z)$ is
\begin{equation}\label{eq:neve}
N_{jk}^{\rm exp} = \sum_{i=1}^{280} \sum_{\alpha=\mu,\tau} \eta_{ijk} \,A_{ijk}^{\nu_\alpha} \,\Phi^{\nu_\mu}_{ik} \,\langle P_{\nu_\mu\to\nu_\alpha}\rangle_{ik} + \{\nu\to\bar{\nu}\}~,
\end{equation} 
where $\langle P_{\nu_\mu\to\nu_\alpha}\rangle_{ik}$ is the average of $\nu_\mu\to\nu_\alpha$ oscillation probability in the $(i,k)$th bins of $(E_\nu,\cos\theta_z)$. In principle, a term containing the contribution of atmospheric electron neutrino flux, $\Phi^{\nu_e}_{ik} \,\langle P_{\nu_e\to\nu_\alpha}\rangle_{ik}$, should be included in Eq.~(\ref{eq:neve}); but since in the high energies $\Phi^{\nu_e}/\Phi^{\nu_\mu}\lesssim 1/30$, we neglect it. In practice, the upper limit of the summation in Eq.~(\ref{eq:neve}) can be replaced by 145, corresponding to $E_\nu\simeq420$~TeV. The effect of higher neutrino energies are $\lesssim1\%$.

The effective areas $A_{ijk}^{\nu_\alpha}$ have been derived from MC simulations and suffer limited statistics. The corresponding uncertainty, $\delta A_{ijk}^{\nu_\alpha}$, is provided by the IceCube collaboration and can be used to estimate the uncertainty on the number of events $\delta N_{jk}^{\rm exp}$ by replacing $A_{ijk}^{\nu_\alpha}\to \delta A_{ijk}^{\nu_\alpha}$ in Eq.~(\ref{eq:neve}). The expected number of events can be confronted with the observed events $N_{jk}^{\rm obs}$ by the following $\chi^2$ function
\begin{equation}\label{eq:chi2}
\chi^2(\vec{\zeta}) = \min_\beta \left\{\sum_{j=1}^{50} \sum_{k=1}^{11} \frac{\left[ N_{jk}^{\rm obs} -\beta N_{jk}^{\rm exp}(\vec{\zeta}) \right]^2}{\left(\sigma_{jk}^{\rm stat}\right)^2 + \left(f_{jk}N_{jk}^{\rm exp}\right)^2} + \frac{\left( 1- \beta \right)^2}{\sigma_\beta^2} \right\}~,
\end{equation}
where $\vec{\zeta}$ is the set of VEP parameters, $\beta$ is the nuisance parameter taking into account the uncertainty on the normalization of atmospheric neutrino flux with $\sigma_\beta=0.24$~\cite{Honda:2006qj} in the pull-term, $\sigma_{jk}^{\rm stat}=\sqrt{N_{jk}^{\rm obs}}$ is the statistical error, and $f_{jk}=\delta N_{jk}^{\rm exp}/ N_{jk}^{\rm exp}$ is the percentage of systematic error in $(j,k)$th bin computed for the standard neutrino oscillation scenario (no VEP). The upper limits of summations in Eq.~(\ref{eq:chi2}) also can be replaced by 25 and 10, respectively corresponding to $E_\mu^{\rm proxy}\simeq31.6$~TeV and $\cos\theta_z=0$, since the contribution of higher energy events are negligible and propagation baseline for events above the horizon at IceCube's site is very short. The constraints on the VEP parameters $\vec{\zeta}$ can be derived from $\Delta\chi^2 (\vec{\zeta})= \chi^2(\vec{\zeta}) - \chi^2_{\rm min} \leq \mathcal{N}$, where $\chi^2_{\rm min}$ is the minimum value of $\chi^2(\vec{\zeta})$ and $\mathcal{N}$ depends on the number of VEP parameters and the desired confidence level of constraint~\cite{Zyla:2020zbs}. The nonzero VEP parameters do not improve the fit to IceCube atmospheric data set and the minimum value $\chi^2_{\rm min}=286.16$ has been obtained for $\vec{\zeta}=0$, which corresponds to the {\it reduced} minimum value $\simeq1.1$ (after dividing by the number of degrees of freedom) that points to a decent fit.

\section{\label{sec:cons}Constraints on VEP parameters}

\begin{figure}[t!]
\centering
\subfloat{
\includegraphics[width=0.49\textwidth]{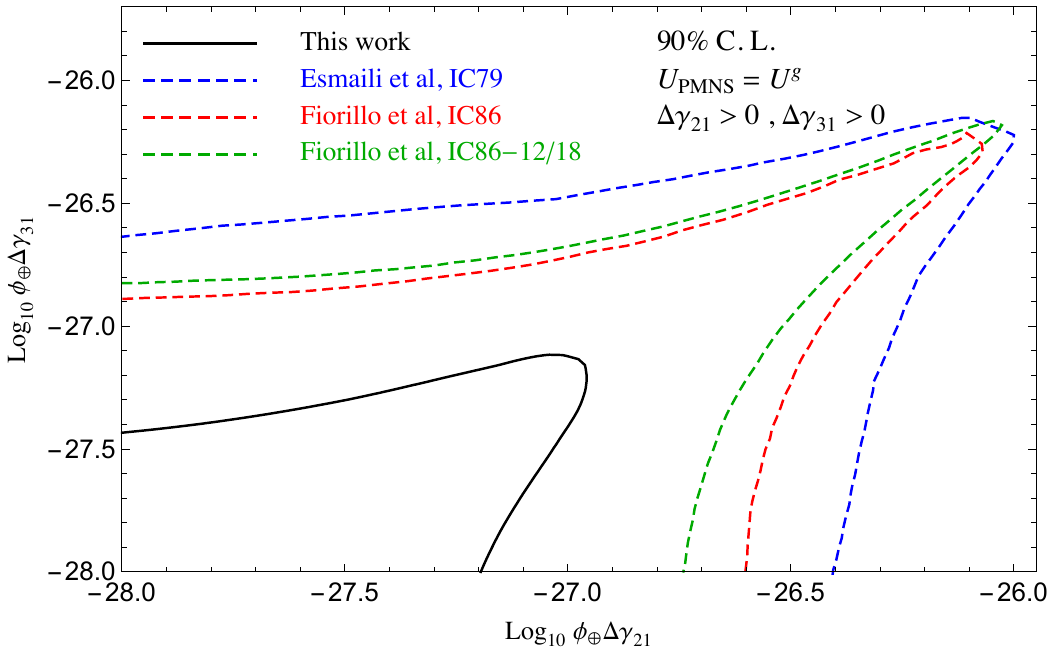}
}
\subfloat{
\includegraphics[width=0.49\textwidth]{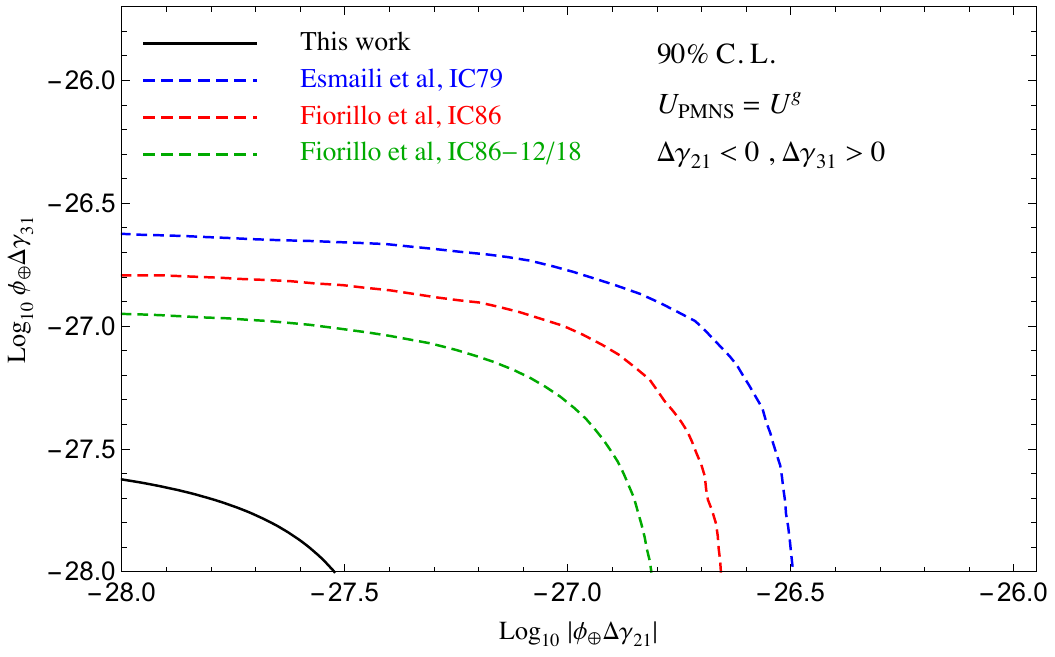}
\label{fig:antinu-gq40-dg13}
}
\caption{\label{fig:uequg}Limit in the plane $(\phi_\oplus\Delta\gamma_{21},\phi_\oplus\Delta\gamma_{31})$ at 90\% C.L., black solid curves, derived from IceCube data set of $\mu$-tracks~\cite{Aartsen:2015rwa}. Previous limits are shown by dashed curves, taken from~\cite{Esmaili:2014ota,Fiorillo:2020gsb}. The left panel is for $\Delta\gamma_{21}, \Delta\gamma_{31} >0$ and the right panel is for $\Delta\gamma_{21}<0, \Delta\gamma_{31} >0$.}
\end{figure}

Equipped with the analysis method explained in section~\ref{sec:data} and the oscillation probabilities of atmospheric neutrinos in the presence of VEP discussed in section~\ref{sec:atm} we can derive constraints on the VEP parameters $(\phi_\oplus\Delta\gamma_{21},\phi_\oplus\Delta\gamma_{31},\gq12,\gq13,\gq23)$. Let us start with the case $U_{\rm PMNS}=U^{\rm g}$ which has been already studied in~\cite{Esmaili:2014ota,Fiorillo:2020gsb}. In this case, where $\theta_{ij}^{\rm g}$ is fixed to the values of $\theta_{ij}$, limits can be derived in the plane of $(\phi_\oplus\Delta\gamma_{21},\phi_\oplus\Delta\gamma_{31})$. Figure~\ref{fig:uequg} shows the obtained 90\% C.L. limit by black solid curve, together with the previous derived limits in~\cite{Esmaili:2014ota,Fiorillo:2020gsb} by dashed curves, for $\Delta\gamma_{21}, \Delta\gamma_{31} >0$ in the left panel, and $\Delta\gamma_{21}<0, \Delta\gamma_{31} >0$ in the right panel. The limits are sensitive just to the relative sign of $\Delta\gamma_{21}$ and $\Delta\gamma_{31}$, so the cases $\Delta\gamma_{21}, \Delta\gamma_{31} <0$ and $\Delta\gamma_{21}>0, \Delta\gamma_{31} <0$ are similar to these figures. The obtained limits in this work are stronger than the previous limits by a factor of few since the detailed information are available for the chosen data set~\cite{Aartsen:2015rwa}, particularly the systematic error from limited statistics in MC simulation of detector. For other data sets this systematic error is not available and should be estimated, leading to less precise $\chi^2$ calculation and weaker limits. Also, all the previous limits (dashed curves in figure~\ref{fig:uequg}) have been obtained from zenith distribution of events and integrating over the energy. The derived limits in this work benefit from both zenith and energy distributions of events and consequently are stronger.

\begin{figure}[t!]
\centering
\subfloat{
\includegraphics[width=0.9\textwidth]{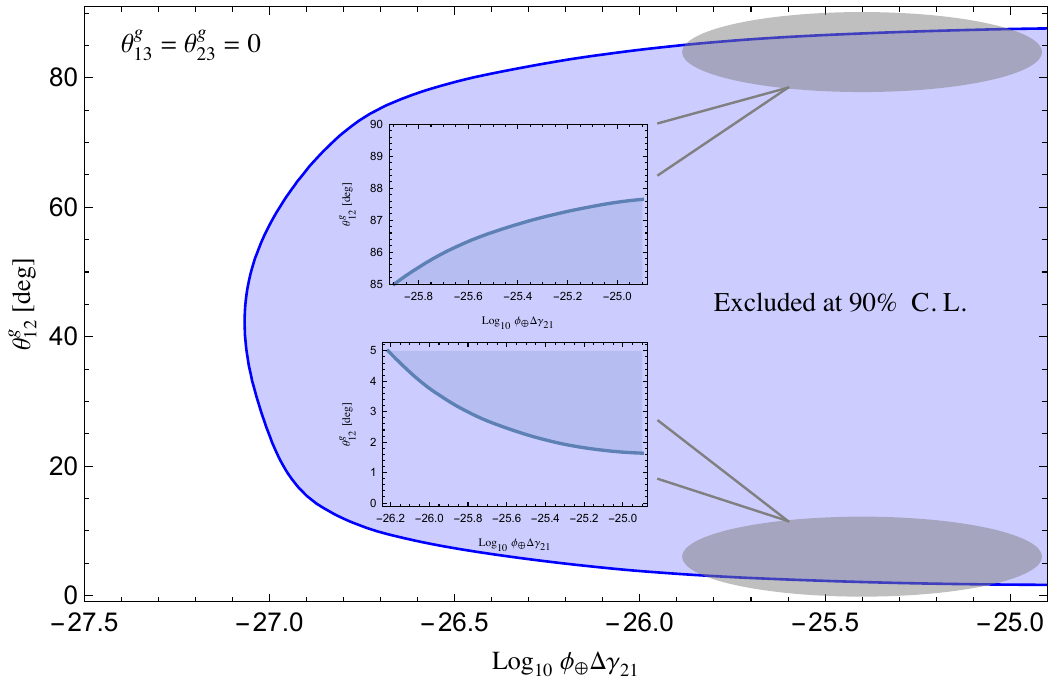}
}
\caption{\label{fig:limgq12}Limit in the plane $(\phi_\oplus\Delta\gamma_{21},\gq12)$ at 90\% C.L., with $\gq13=\gq23=0$, from atmospheric neutrino data of IceCube~\cite{Aartsen:2015rwa}. The insets show the enlarged regions of $0\leq\gq12\leq5^\circ$ and $85^\circ\leq\gq12\leq90^\circ$. }
\end{figure}

Relaxing the condition $U_{\rm PMNS}=U^{\rm g}$, we can set limits on the angles $\theta_{ij}^{\rm g}$ which parameterize $U^{\rm g}$. Figure~\ref{fig:limgq12} shows the bound in the plane $(\phi_\oplus\Delta\gamma_{21},\gq12)$ at 90\% C.L., assuming $\gq13=\gq23=0$. As we discussed in section~\ref{sec:gq12}, the $\Delta\gamma_{31}$ cannot be constrained in this case. For a better visualization, in the insets we show the enlarged regions of $0\leq\gq12\leq5^\circ$ and $85^\circ\leq\gq12\leq90^\circ$. The slight asymmetry of the exclusion curve with respect to $\gq12=\pi/4$ originates from the fact that for the second octant, $\gq12 > \pi/4$, the resonance occurs in anti-neutrino channel where the atmospheric flux is smaller by a factor of $\sim2$ in comparison with neutrino flux in $\sim$~TeV energies. This asymmetry can be seen better in the insets: $\phi_\oplus\Delta\gamma_{21}>10^{-25}$ is rejected at 90\% C.L. for $\gq12<87.5^\circ$ in the second octant, while it can be rejected for $\gq12>1.7^\circ$ in the first octant. The asymmetry becomes less pronounced by increasing $\gq12$ (or increasing $\pi/4-\gq12$) since the fast oscillations in high energy are present in both neutrino and anti-neutrino channels. In figure~\ref{fig:limgq12} we assumed $\Delta\gamma_{21}>0$. The limit for $\Delta\gamma_{21}<0$ will be reflection of the curve in figure~\ref{fig:limgq12} with respect to $\gq12=\pi/4$ since the resonances for $\Delta\gamma_{21}<0$ are in neutrino and anti-neutrino channel when $\gq12$ is in the second and first octant, respectively.

As we discussed in section~\ref{sec:gq13}, for $\gq13\neq0$ the oscillation probabilities in $e-\tau$ sector will be modified. In principle, deviations in $P_{\nu_e\to\nu_e(\nu_\tau)}$ and $P_{\bar{\nu}_e\to\bar{\nu}_e(\bar{\nu}_\tau)}$ can alter the energy and zenith angle distributions of IceCube $\mu$-track events, since the leptonic decays of $\tau^\pm$, from the charged-current interaction of tau-neutrinos, produce $\mu^\pm$. However, this effect is very small for two reasons: small branching ratio of the leptonic decay of tau $\sim17\%$, and small electron-neutrino atmospheric flux ($\Phi^{\nu_e}/\Phi^{\nu_\mu}\sim1/30$ at $E_\nu \gtrsim 100$~GeV). We have checked that by the current sensitivity of IceCube detector and the collected statistics, no limits can be derived on $\Delta\gamma_{31}$ for $\gq13\neq0$, from $\mu$-track data set. IceCube has measured the atmospheric $\nu_e+\bar{\nu}_e$ flux at energies $\gtrsim100$~GeV~\cite{Aartsen:2015xup}. However, this measurement uses the cascade (or shower) events which can be produced through both electron and tau neutrino charged and neutral current interactions. Thus, deviations in $P_{\nu_e\to\nu_e(\nu_\tau)}$ and $P_{\bar{\nu}_e\to\bar{\nu}_e(\bar{\nu}_\tau)}$, which occur for $\gq13\neq0$, cannot be demonstrated in this measurement.

\begin{figure}[t!]
\centering
\subfloat{
\includegraphics[width=0.7\textwidth]{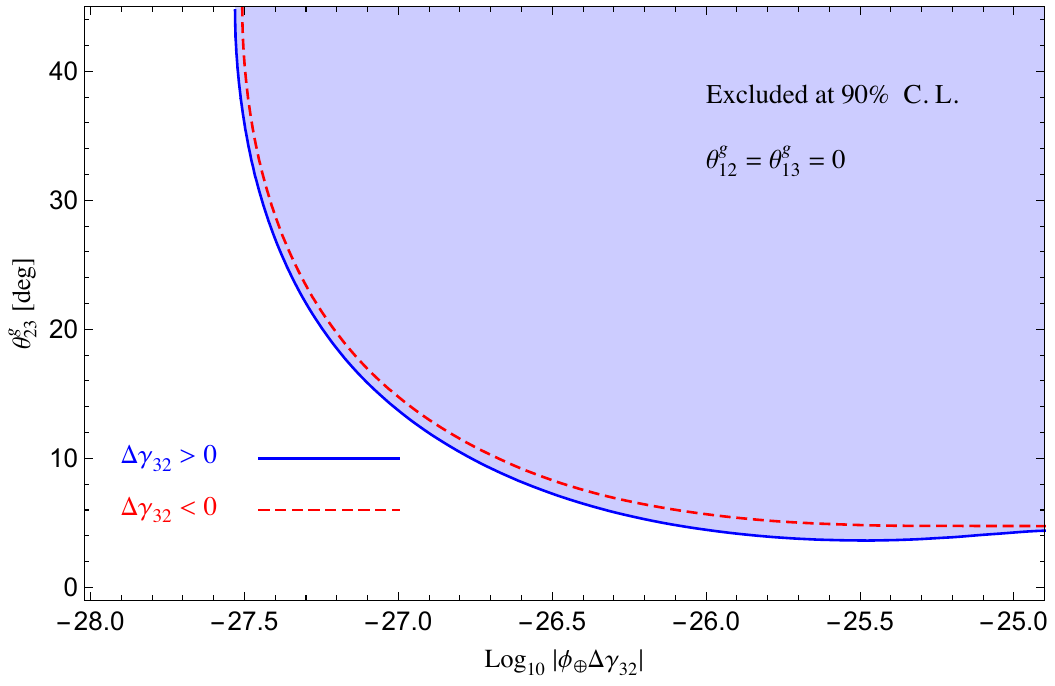}
}
\caption{\label{fig:limgq23}Limit in the plane $(\phi_\oplus\Delta\gamma_{32},\gq23)$ at 90\% C.L., with $\gq12=\gq13=0$, from atmospheric neutrino data of IceCube~\cite{Aartsen:2015rwa}. The solid (dashed) curve corresponds to $\Delta\gamma_{32}>0$ ($\Delta\gamma_{32}<0$).}
\end{figure}

Finally, in figure~\ref{fig:limgq23} we show the constraint derived in $(\phi_\oplus\Delta\gamma_{32},\gq23)$ at 90\% C.L. from atmospheric neutrinos. As discussed in section~\ref{sec:gq23}, when $\gq23\neq0$ the induced oscillation in the $\mu-\tau$ sector can be explained by an effective mass-squared difference $\Delta m^2_{\rm eff}=2E_\nu^2\phi_\oplus(\Delta\gamma_{31}-\Delta\gamma_{21})=2E_\nu^2\phi_\oplus\Delta\gamma_{32}$. The sign of $\Delta\gamma_{32}$ is important just in the region where the VEP-induced and $\Delta m^2_{31}$-induced oscillations interfere, that is $E_\nu\lesssim100$~GeV, if $\Delta\gamma_{32}$ has the appropriate value such that VEP-induced oscillations extend down to this region. However, since the effective area is much smaller in this region, we get practically the same constraints for both signs of $\Delta\gamma_{32}$ (the solid and dashed curves in figure~\ref{fig:limgq23}). The exclusion curves for $\pi/4\leq\gq23\leq\pi/2$ are the reflections of curves in figure~\ref{fig:limgq23} with respect to $\gq23=\pi/4$. From figure~\ref{fig:limgq23}, $\gq23\gtrsim4^\circ$ (or $\pi/2-\gq23\gtrsim4^\circ$) can be excluded at 90\% C.L. for $\phi_\oplus\Delta\gamma_{32}\gtrsim10^{-26}$.

\begin{figure}[t!]
\centering
\subfloat{
\includegraphics[width=0.7\textwidth]{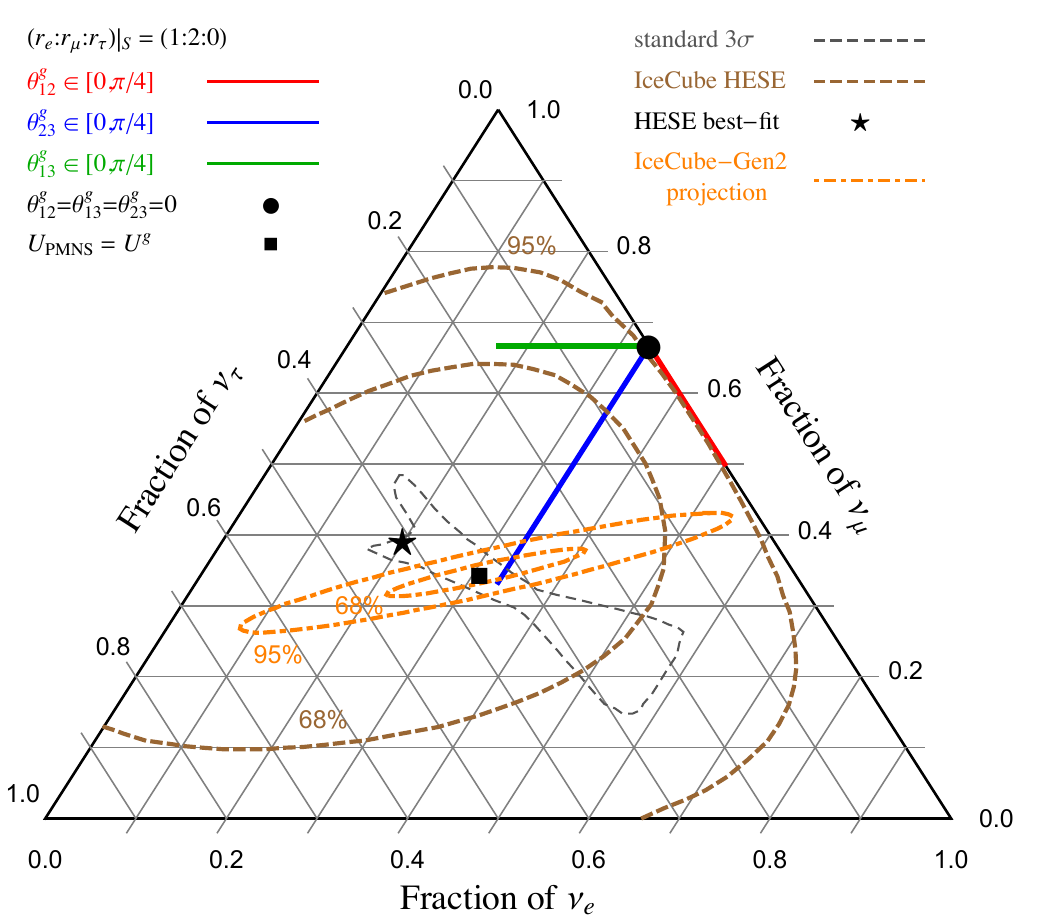}
}
\caption{\label{fig:terpion}Ternary plot of the flavor content of cosmic neutrinos at Earth for $(r_e:r_\mu:r_\tau)|_{\rm S}=1:2:0$ from pion decay chain at source. The $\sqcdot$ and $\bullet$ show the flavor content for $U^{\rm g}=U_{\rm PMNS}$ and $U^{\rm g}=I$, respectively. The red, blue and green lines starting from $\bullet$ show the flavor contents for non-zero $\gq12$, $\gq23$ and $\gq13$, respectively, varying in $[0,\pi/4]$. The current allowed regions by IceCube HESE data set~\cite{Abbasi:2020zmr} and projected sensitivity of IceCube-Gen2~\cite{Bustamante:2019sdb} (at 68\% and 95\% C.L.) are shown by brown dashed and orange dot-dashed curves, respectively.}
\end{figure}

\begin{figure}[t!]
\centering
\subfloat{
\includegraphics[width=0.7\textwidth]{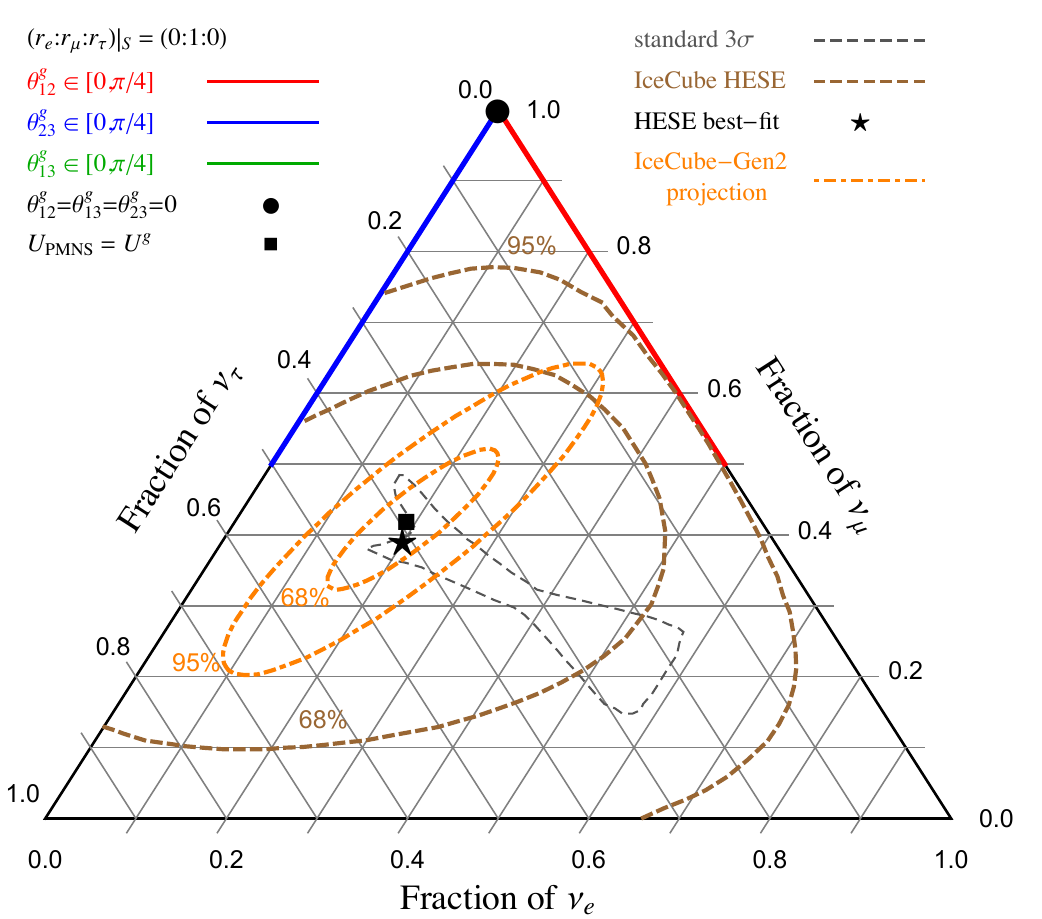}
}
\caption{\label{fig:terdamp}The same as figure~\ref{fig:terpion} but for muon damped scenario of flavor ratio at the source $(r_e:r_\mu:r_\tau)|_{\rm S}=0:1:0$.}
\end{figure}

Knowing the constraints derived from atmospheric neutrinos on the parameters of $U^{\rm g}$, that are $(\gq12,\gq13,\gq23)$, let us turn to the constraints that can be derived from the flavor measurement of cosmic neutrinos. These constraints are summarized in figure~\ref{fig:terpion} which shows the ternary plot of the flavor content of cosmic neutrinos at Earth assuming the flavor content $(r_e:r_\mu:r_\tau)|_{\rm S}=1:2:0$ from pion decay chain at source. The current IceCube best-fit point from HESE data set~\cite{Abbasi:2020zmr} is shown by $\star$ symbol, and the 68\% and 95\% C.L. limits are shown by dashed brown curves. The dashed gray contour shows the expected flavor content allowing the standard mixing angles, $\theta_{ij}$, vary within their $3\sigma$ ranges. The $\sqcdot$ shows the expected flavor content when $U_{\rm PMNS}=U^{\rm g}$, that is when the flavor content of cosmic neutrinos changes from source to Earth according to the expectation from standard scenario (no VEP effect on cosmic neutrinos). The $\theta^{\rm g}_{ij}=0$ is shown by $\bullet$. Notice that the flavor content shown by $\bullet$ is expected for any $\phi_{\rm IG}\Delta\gamma_{ij}\gtrsim6\times10^{-32}$, see section~\ref{sec:cosmic}. The departure of VEP mixing angles from zero are shown by the colored straight lines emerging from $\bullet$: the red, green and blue lines correspond to $\gq12\neq0$, $\gq13\neq0$ and $\gq23\neq0$, respectively\footnote{The legends in figures~\ref{fig:terpion} and \ref{fig:terdamp} indicate $[0,\pi/4]$ for the range of $\theta_{ij}^{\rm g}$. Variation in the range $[\pi/4,\pi/2]$ leads to the same lines since the oscillation probabilities of cosmic neutrinos depend on $|U_{\alpha i}^{\rm g}|^2$.}. As can be seen, by the current HESE data, any value of $\gq12$ and $\gq13$ can be excluded by more than 68\% C.L., and the bound $\gq23>15.5^\circ (4^\circ)$ can be set at 68\% (95\%) C.L. (the intersection point of blue line and brown dashed curves in figure~\ref{fig:terpion}). The orange dot-dashed contours show the 68\% and 90\% C.L. constraints that can be achieved by the IceCube-Gen2~\cite{Aartsen:2020fgd}, taken from~\cite{Bustamante:2019sdb}. The segment of blue line inside the 68\% (95\%) C.L. contour of IceCube-Gen2 corresponds to $\gq23>35^\circ (32.5^\circ)$.

Figure~\ref{fig:terdamp} shows the ternary plot of flavor content at Earth for muon damped scenario at the source $(r_e:r_\mu:r_\tau)|_{\rm S}=0:1:0$, with the same symbols and legends as Figure~\ref{fig:terpion}. Clearly, for muon damped scenario the flavor content of cosmic neutrinos at Earth does not depend on $\gq13$, and so varying $\gq13\in[0,\pi/4]$ the flavor content at Earth remains equal to $0:1:0$. The red line, corresponding to $\gq12 \in [0,\pi/4]$ is excluded at more than 95\% C.L. by the current HESE data, while the limit $\gq23>35^\circ (23^\circ)$ at 68\% (95\%) C.L. can be set from the blue line.      

\section{\label{sec:conc}Discussions and Conclusions}

The coupling of neutrinos to gravity is diagonal in the gravitational basis $\nu^{\rm g}_i$ which can be different from both mass basis $\nu_i$ and flavor basis $\nu_\alpha$, where $i=1,2,3$ and $\alpha=e,\mu,\tau$.  Consequently, similar to the PMNS unitary matrix which transforms the mass basis to flavor basis, the transformation of gravitational basis to flavor basis can be casted into the $3\times3$ unitary matrix $U^{\rm g}$, parameterized by three angles $(\gq12,\gq13,\gq23)$ (assuming vanishing phases since their effect is marginal). In the presence of VEP, the gravitational coupling of $\nu_i^{\rm g}$ is $G_{N,i}=\gamma_iG_N$, where non-equality of $\gamma_i$ factors induces VEP. While neutrino flavor oscillation is not sensitive to the values of $\gamma_i$, non-zero $\Delta\gamma_{ij}=\gamma_i-\gamma_j$ modifies the oscillation probabilities. In the following we summarize and discuss the constraints that we derived from atmospheric and cosmic neutrino data, and the synergy of them, on the five parameters $(\theta_{ij}^{\rm g},\Delta\gamma_{ij})$ of VEP. Clearly, when $\gamma_i=\gamma_j$ (for $i\neq j$) and so $\Delta\gamma_{ij}=0$, the corresponding angle $\theta_{ij}^{\rm g}$ is redundant and the number of VEP parameters reduce to three.  

Measurements of atmospheric neutrino oscillation cannot constrain $\gq13$ since $\gq13\neq0$ induces $e\leftrightarrow\tau$ oscillations while the flux of atmospheric electron and tau neutrinos are small in the high energy and neutrino telescopes, such as IceCube, cannot distinguish these two flavors (at least up to few hundreds of TeV where the atmospheric neutrino flux is negligible). On the other hand, measurements of the cosmic neutrino's flavor content can constrain $\gq13$ for any value $\phi_{\rm IG}\Delta\gamma_{31}\gtrsim10^{-32}$. By the current data any $\gq13$ is excluded at $\gtrsim95\%$ C.L. for muon damped sources and at $\gtrsim68\%$ C.L. for pion decay chain scenario. In the future IceCube-Gen2 can exclude $\gq13$ at higher confidence levels.

Both the $\gq12$ and $\gq23$ angles can be constrained by atmospheric neutrino data. From these data, $\gq12\gtrsim2^\circ (25^\circ)$, or $\gq12\lesssim88^\circ (57^\circ)$ in the second octant, is excluded for $\phi_\oplus\Delta\gamma_{21}\gtrsim10^{-25} (10^{-27})$. From the measurement of cosmic neutrino's flavor content, for both the damped muon and pion decay chain scenarios, any $\gq12$ for $\phi_{\rm IG}\Delta\gamma_{21}\gtrsim10^{-32}$ can be excluded at $\gtrsim95\%$ C.L. by the current data, and IceCube-Gen2 will improve the exclusion significance notably. For $\gq23$ the synergy of cosmic and atmospheric neutrino data is  more crucial. While the flavor content measurement sets the bound $\gq23\gtrsim15^\circ (4^\circ)$ for pion decay chain sources and $\gtrsim35^\circ (23^\circ)$ for muon damped sources, at 68\% (95\%) C.L. and for $\phi_{\rm IG}\Delta\gamma_{32}\gtrsim10^{-32}$, the atmospheric neutrino data exclude $\gq23>4^\circ$ at 90\% C.L. for $\phi_\oplus\Delta\gamma_{32}\gtrsim10^{-26}$. Future IceCube-Gen2 data can exclude any $\gq23$ at $\gtrsim95\%$ C.L. for muon damped sources and exclude $\gq23<32.5^\circ$ at 95\% C.L. for pion decay chain sources.   

From the constraints derived in this article, and noticing that $\phi_{\rm IG}\simeq\phi_\oplus$, we conclude that all the extended parameter space of VEP in neutrino sector, that is all the possibilities for $\theta_{ij}^{\rm g}$, are already excluded by the synergy of cosmic and atmospheric neutrino data for $\phi_\oplus\Delta\gamma_{ij}\gtrsim10^{-26}$. For smaller $\phi_\oplus\Delta\gamma_{ij}$ the atmospheric neutrino data lose their constraining power, such that for $\phi_\oplus\Delta\gamma_{ij}\lesssim10^{-28}$ no constraints can be derived on $\theta_{ij}^{\rm g}$. For this range of $\phi_\oplus\Delta\gamma_{ij}$ the measurement of cosmic neutrino's flavor content is crucial and IceCube-Gen2 can exclude any $\theta_{ij}^{\rm g}$ for muon damped sources and any $(\gq12,\gq13)$ and $\gq23\gtrsim32^\circ$ for pion decay chain sources.

\begin{acknowledgments}
The author thanks John David Rogers Computing Center (CCJDR) in the Institute of Physics ``Gleb Wataghin'', University of Campinas, for providing computing resources.
\end{acknowledgments}


\bibliographystyle{JHEP}
\bibliography{refs}

\end{document}